\documentclass[prd,reprint,showpacs,showkeys]{revtex4}
\usepackage{eurosym}
\usepackage{amsfonts}
\usepackage{amssymb}
\usepackage{caption}
\usepackage{amsmath}
\usepackage{graphicx}

\begin{document}

\title{On a Particular Thin-shell Wormhole }
\author{A. \"{O}vg\"{u}n}
\email{ali.ovgun@emu.edu.tr}
\author{I. Sakalli}
\email{izzet.sakalli@emu.edu.tr}
\affiliation{Physics Department , Eastern Mediterranean University, Famagusta, Northern
Cyprus, Mersin 10, Turkey}
\date{\today }

\begin{abstract}
In this paper, using a black hole with scalar hair, we construct a scalar
thin-shell wormhole (TSW) in 2+1 dimensions by applying the Visser's famous
cut and paste technique. The surface stress, which is concentrated at the
wormhole throat is determined by using the Darmois-Israel formalism. By
employing the various gas models, we study the stability analysis of the
TSW. The region of stability is changed by tuning the parameters of $\ "l"$
and $"u"$. It is observed that the obtained particular TSW originated from
the black hole with scalar hair could be more stable with particular $l$
parameter, however it still needs exotic matter..
\end{abstract}

\maketitle

\section{Introduction}

In 1988 Morris and Thorne \cite{morris} devised the traversable wormholes,
which are the solutions of the Einstein's equations of gravitation. They are
cosmic shortcuts that connect two points of the Universe by a throat-like
geometry. However, they violate one or more of the so-called energy
conditions [weak energy condition (WEC), null energy condition (NEC), and
strong energy condition (SEC)]; see for instance Refs. \cite%
{yurtsever,visser1,visser2}. Because of this fact, most physicists agree
that wormholes require exotic matter -- a kind of antigravity -- to keep
their throat (the narrowest point) open \cite{witt}. Conversely, some of the
physicists studying in this subject claimed that wormholes like the TSW can
be supported by normal matter \cite{nor1,nor3}.

Firstly, Visser \cite{visser3} proposed the method of how the TSWs can be
constructed via the Israel's junction conditions \cite{israel}. It is shown
that the amount of exotic matter \cite{visser4} around the throat can be
minimized with a suitable choice of the geometry of the wormhole. Following
the Visser's prescription, today there are many studies in the literature
focused on the build of the TSWs described in the arbitrary (lower/higher)
dimensions (see for instance \cite%
{nor2,lemos1,camera,banerjee2,banerjee7,banerjee3,banerjee6,lobo,kuf,ali1,darabi,Kas,sharif1,sharif2,sharif3,sime1,sharif4,sharif5,eid,banerjee8,banerjee,banerjee9,banerjee10}%
). In this paper, we consider the scalar hair black hole (SHBH) in 2+1
dimensions that is the solution to the Einstein-Maxwell theory with
self-interacting scalar field described by the Liouville potential $V\left(
\phi \right) $ \cite{MH}. Then, using the standard procedure of the cut and
paste technique we construct TSW, and test its stability for the different
physical gas states.

\textit{Our main motivation in constructing a thinshell wormhole is to
minimize the exotic matter, which is in general the main source for
supporting the throat. In this paper, we focus on the \ stability of the
SHBH spacetime in 2+1 dimensions inasmuch as this black hole depends on two
variables and by fixing them, it is possible to reach a stable solutions.}

The paper is organized as follows: In Sec. 2, we first give a brief
introduction about the SHBH described within 2+1 dimensional geometry. In
Sec. 3, we firstly setup TSW, and then apply the various gas models to the
equation of state (EoS) for testing its stability. The paper ends with our
conclusions in Sect. 4.

\section{SHBH Spacetime}

In this section, we shall briefly overview the SHBH \cite{MH}. The following
action describes the Einstein-Maxwell gravity that is minimally coupled to a
scalar field $\phi $ \ 
\begin{equation}
S=\int d^{3}r\sqrt{-g}\left( R-2\partial _{\mu }\phi \partial ^{\mu }\phi
-F^{2}-V\left( \phi \right) \right) ,  \label{eq1}
\end{equation}%
where $R$ denotes the Ricci scalar, $F=F_{\mu \nu }F^{\mu \nu }$ is the
Maxwell invariant, and $V\left( \phi \right) $ stands for the scalar ($\phi$%
) potential.SHBH is the solution to the action (1), which was found by \cite%
{MH} as follows 
\begin{equation}
ds^{2}=-f(r)dt^{2}+\frac{4r^{2}dr^{2}}{f(r)}+r^{2}d\theta ^{2},  \label{eq2}
\end{equation}%
where the metric function is given by 
\begin{equation}
f(r)=\frac{r^{2}}{l{^{2}}}-ur.  \label{eq3}
\end{equation}%
Here $u$ and $l$ are constants, and event horizon of the BH (2) is located
at $r_{h}=u\ell ^{2}$. It is clear that this BH possesses a
non-asymptotically flat geometry. Metric (2) can alternatively be rewritten
in the following form 
\begin{equation}
ds^{2}=-\frac{r}{\ell ^{2}}\left( r-r_{h}\right) dt^{2}+\frac{4r\ell
^{2}dr^{2}}{\left( r-r_{h}\right) }+r^{2}d\theta ^{2}.
\end{equation}%
The singularity located at $r=0$ can be best seen by checking the Ricci and
Kretschmann scalars: 
\begin{equation}
R=-\frac{2r+r_{h}}{4r^{3}\ell ^{2}},
\end{equation}%
\begin{equation}
K=\frac{4r^{2}-4r_{h}r+3r_{h}^{2}}{16r^{6}\ell ^{4}}.
\end{equation}%
The scalar field and potential are respectively given by \cite{MH}

\begin{equation}
\phi =\frac{\ln r}{\sqrt{2}},
\end{equation}

\begin{equation}
V\left( \phi \right) =\frac{\lambda _{1}+\lambda _{2}}{r^{2}},
\end{equation}%
in which $\lambda _{1,2}$ are constants. The corresponding Hawking
temperature of the BH (see, for example, \cite{sakalli0}) is as follows 
\begin{equation}
T_{H}=\frac{1}{4\pi }\left. \frac{\partial {f}}{\partial {r}}\right\vert
_{r=r_{h}}=\frac{1}{8\pi \ell ^{2}},
\end{equation}%
which is constant. Having a radiation with constant temperature is the
well-known isothermal process. It is worth noting that Hawking radiation of
the linear dilaton black holes exhibits similar isothermal behavior \cite%
{Clem1,sakalli0,sakalli1,clem2,sakalli2,ali2,ali22,sakalli11,sakalli12}.

\section{Stability of TSW}

In this section, we take two identical copies of the SHBHs with $(r\geq a)$: 
\begin{equation*}
M^{\pm }=(x|r\geq 0),
\end{equation*}%
and the manifolds are bounded by hypersurfaces $M^{+}$ and $M^{-}$, to get
the single manifold $M=M^{+}+M^{-}$, we glue them together at the surface of
the junction 
\begin{equation*}
\Sigma ^{\pm }=(x|r=a).
\end{equation*}%
where the boundaries $\Sigma $ are given . On the shell, the spacetime can
be chosen to be%
\begin{equation}
ds^{2}=-d\tau ^{2}+a(\tau )^{2}d\theta ^{2},
\end{equation}

where $\tau $ represents the proper time \cite{ali1}. Setting coordinates $%
\xi ^{i}=(\tau ,\theta )$, the extrinsic curvature formula connecting the
two sides of the shell is simply given by \cite{sime1} 
\begin{equation}
K_{ij}^{\pm }=-n_{\gamma }^{\pm }\left( \frac{\partial ^{2}x^{\gamma }}{%
\partial \xi ^{i}\partial \xi ^{j}}+\Gamma _{\alpha \beta }^{\gamma }\frac{%
\partial x^{\alpha }}{\partial \xi ^{i}}\frac{\partial x^{\beta }}{\partial
\xi ^{j}}\right) ,  \label{extcur}
\end{equation}%
where the unit normals ($n^{\gamma }n_{\gamma }=1)$ are 
\begin{equation}
n_{\gamma }^{\pm }=\pm \left\vert g^{\alpha \beta }\frac{\partial H}{%
\partial x^{\alpha }}\frac{\partial H}{\partial x^{\beta }}\right\vert
^{-1/2}\frac{\partial H}{\partial x^{\gamma }},  \label{normgen}
\end{equation}%
with $H(r)=r-a(\tau )$. Using the metric functions, the non zero components
of $n_{\gamma }^{\pm }$ become 
\begin{equation}
n_{t}=\mp 2a\dot{a},
\end{equation}%
\begin{equation}
n_{r}=\pm 2\sqrt{\frac{al^{2}(4\dot{a}^{2}l^{2}a-l^{2}u+a)}{(l^{2}u-a)}},
\end{equation}%
where the dot over a quantity denotes the derivative with respect to $\tau $%
. Then, the non-zero extrinsic curvature (\ref{extcur}) components yield 
\begin{equation}
K_{\tau \tau }^{\pm }=\mp \frac{\sqrt{-al^{2}}(8\dot{a}^{2}l^{2}a+8\ddot{a}%
l^{2}a^{2}-l^{2}u+2a)}{4a^{2}l^{2}\sqrt{-4\dot{a}^{2}l^{2}a-l^{2}u+a}},
\end{equation}%
\begin{equation}
K_{\theta \theta }^{\pm }=\pm \frac{1}{2a^{\frac{3}{2}}l}\sqrt{4\dot{a}%
^{2}l^{2}a-l^{2}u+a}.
\end{equation}

Since $K_{ij}$ is not continuous around the shell ($H$) \cite{sime1}, we use
the Lanczos equation \cite{lan1,darmois,lake}:

\begin{equation}
S_{ij}=-\frac{1}{8\pi }\left( [K_{ij}]-[K]g_{ij}\right) .  \label{ee}
\end{equation}

where $K$ is the trace of $K_{ij}$, $[K_{ij}]=K_{ij}^{+}-K_{ij}^{-},$ and $%
S_{ij}$ is stress energy-momentum tensor at the junction which is given in
general by \cite{sime1,nor2} 
\begin{equation}
S_{j}^{i}=diag(\sigma ,-p),  \label{enerji}
\end{equation}%
where $p$ is surface pressure, and $\sigma $ is surface energy density. Due
to the circular symmetry, we have 
\begin{equation}
K_{j}^{i}=\left( {%
\begin{array}{cc}
K_{\tau }^{\tau } & 0 \\ 
0 & K_{\theta }^{\theta } \\ 
& 
\end{array}%
}\right) .
\end{equation}%
Thus, from Eq.s (\ref{enerji}) and (\ref{ee}) one obtains the surface
pressure and surface energy density \cite{sime1}.

Using the cut and paste technique, we can demount the interior regions $r<a$
of the geometry (10), and links its exterior parts. However, there exists a
bounce (deduced from the extrinsic curvature components at the surface $r=a$%
) that is related with the energy density and pressure: 
\begin{equation}
\sigma =-\frac{1}{8\pi a^{\frac{3}{2}}l}\sqrt{4\dot{a}^{2}l^{2}a-l^{2}u+a},
\label{ed}
\end{equation}%
\begin{equation}
p=\frac{1}{16\pi a^{\frac{3}{2}}l}\frac{\left( 8\dot{a}^{2}l^{2}a+8\ddot{a}%
l^{2}a^{2}-l^{2}u+2a\right) }{\sqrt{4\dot{a}^{2}l^{2}a-l^{2}u+a}}.
\label{pre}
\end{equation}

Consequently, the energy and pressure quantities in a static case ($a=a_{0}$%
) become 
\begin{equation}
\sigma _{0}=-\frac{1}{8\pi a_{0}^{\frac{3}{2}}l}\sqrt{-l^{2}u+a_{0}},
\end{equation}%
\begin{equation}
p_{0}=\frac{1}{16\pi a_{0}^{\frac{3}{2}}l}\frac{\left( -l^{2}u+2a_{0}\right) 
}{\sqrt{-l^{2}u+a_{0}}}.
\end{equation}

Once $\sigma \geq 0$ and $\sigma +p\geq 0$ hold, then WEC is satisfied.
Besides, $\sigma +p\geq 0$ is the condition of NEC. Furthermore, SEC is
conditional on $\sigma +p\geq 0$ and $\sigma +2p\geq 0$. It is obvious from
Eq. (24) that negative energy density violates the WEC, and consequently we
are in need of the exotic matter for constructing TSW. We note that the
total matter supporting the wormhole is given by \cite{nandi}

\begin{equation}
\Omega _{\sigma }=\int_{0}^{2\pi }\left. [\rho \sqrt{-g}]\right\vert
_{r=a_{0}}d\phi =2\pi a_{0}\sigma (a_{0})=-\frac{1}{4a_{0}^{\frac{1}{2}}|l|}%
\sqrt{-l^{2}u+a_{0}}.
\end{equation}

Stability of such a wormhole is investigated through a linear perturbation
in which the equation of state is given by 
\begin{equation}
p=\psi (\sigma ),
\end{equation}%
where $\psi (\sigma )$ is an arbitrary function of $\sigma $. The energy
conservation equation is introduced as follows \cite{sime1} 
\begin{equation}
S_{j;i}^{i}=-T_{\alpha \beta }\frac{\partial x^{\alpha }}{\partial \xi ^{j}}%
n^{\beta },
\end{equation}%
where $T_{\alpha \beta }$ is the bulk energy-momentum tensor. Eq. (28) can
thus be rewritten in terms of the pressure and energy density: 
\begin{equation}
\frac{d}{d\tau }\left( \sigma a\right) +\psi \frac{da}{d\tau }=-\dot{a}%
\sigma .
\end{equation}%
From above equation, one reads 
\begin{equation}
\sigma ^{\prime }=-\frac{1}{a}(2\sigma +\psi ),
\end{equation}%
and its second derivative yields%
\begin{equation}
\sigma ^{\prime \prime }=\frac{2}{a^{2}}(\tilde{\psi}+3)(\sigma +\frac{\psi 
}{2}).
\end{equation}

where prime and tilde symbols denote derivative with respect to $a$ and $%
\sigma $, respectively. The equation of motion for the shell is in general
given by 
\begin{equation}
\dot{a}^{2}+V=0,
\end{equation}%
where the effective potential $V$ is found from Eq. (22) as 
\begin{equation}
V=\frac{1}{4l^{2}}-\frac{u}{4a}-16a^{2}\sigma ^{2}\pi ^{2}.
\end{equation}%
In fact, Eq. (32) is nothing but the equation of the oscillatory motion in
which the stability around the equilibrium point $a=a_{0}$ is conditional on 
$V^{\prime \prime }(a_{0})\geq 0$. Using Eqs. (30) and (31), we finally
obtain 
\begin{equation}
V^{\prime \prime }=\left. -\frac{1}{2a^{3}}\left[ 64\pi ^{2}a^{5}\left(
\left( \sigma \sigma ^{\prime }\right) ^{\prime }+4\sigma ^{\prime }\frac{%
\sigma }{a}+\frac{\sigma ^{2}}{a^{2}}\right) +u\right] \right\vert
_{a=a_{0}},
\end{equation}%
or equivalently, 
\begin{equation}
V^{\prime \prime }=\left. \frac{1}{2a^{3}}\{-64\pi ^{2}a^{3}\left[ (2\psi
^{\prime }+3)\sigma ^{2}+\psi (\psi ^{\prime }+3)\sigma +\psi ^{2}\right]
-u\}\right\vert _{a=a_{0}}.
\end{equation}

\textit{The equation of motion of the throat, for a small perturbation
becomes \cite{varela,bro1,igor} }%
\begin{equation*}
\dot{a}^{2}+\frac{V^{\prime \prime }(a_{0})}{2}(a-a_{0})^{2}=0.
\end{equation*}

\textit{Noted that for the condition of }$V^{\prime \prime }(a_{0})\geq 0$%
\textit{\ , TSW is stable where the motion of the throat is oscillatory with
angular frequency }$\omega =\sqrt{\frac{V^{\prime \prime }(a_{0})}{2}}$%
\textit{.}

\section{Some Models of EoS Supporting TSW}

In this section, we use particular gas models (linear barotropic gas (LBG) 
\cite{kuf2,varela}, chaplygin gas (CG) \cite{cg,cg1}, generalized chaplygin
gas (GCG) \cite{gcg} and logarithmic gas (LogG) \cite{ali1}) to explore the
stability of TSW.

\subsection{Stability analysis of TSW via the LBG}

The equation of state of LBG \cite{kuf2,varela} is given by 
\begin{equation}
\psi =\varepsilon _{0}\sigma ,
\end{equation}%
and hence 
\begin{equation}
\psi ^{\prime }(\sigma _{0})=\varepsilon _{0},
\end{equation}%
where $\varepsilon _{0}$ is a constant parameter. By changing the values of $%
l$ and $u$ in Eq. (35), we illustrate the stability regions for TSW, in
terms of $\varepsilon _{0}$ and $a_{0}$, as depicted in Fig. 1.

\begin{figure}[h!]
\caption{Stability Regions via the LBG}\centering
\includegraphics[width=0.40\textwidth]{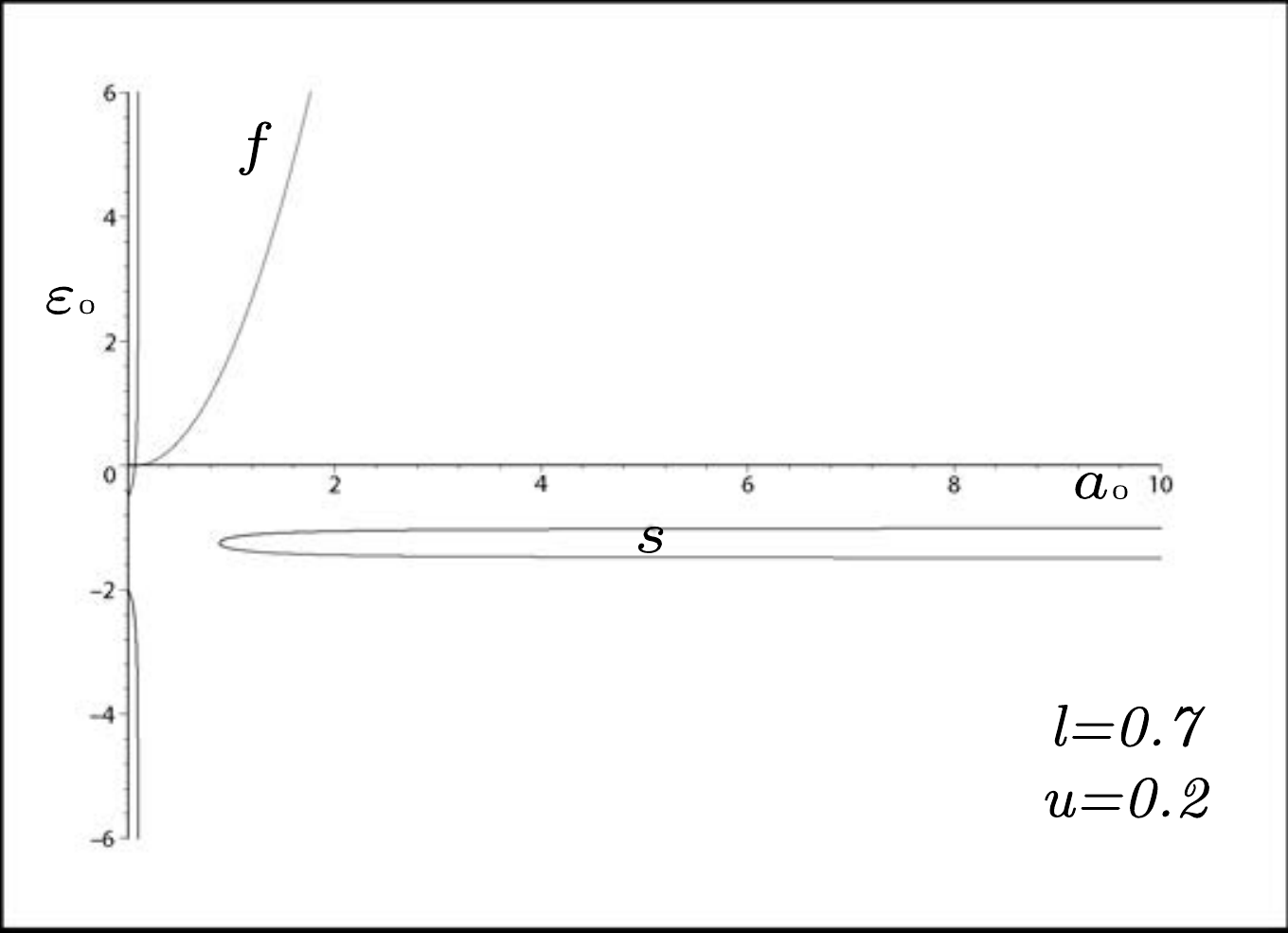} %
\includegraphics[width=0.40\textwidth]{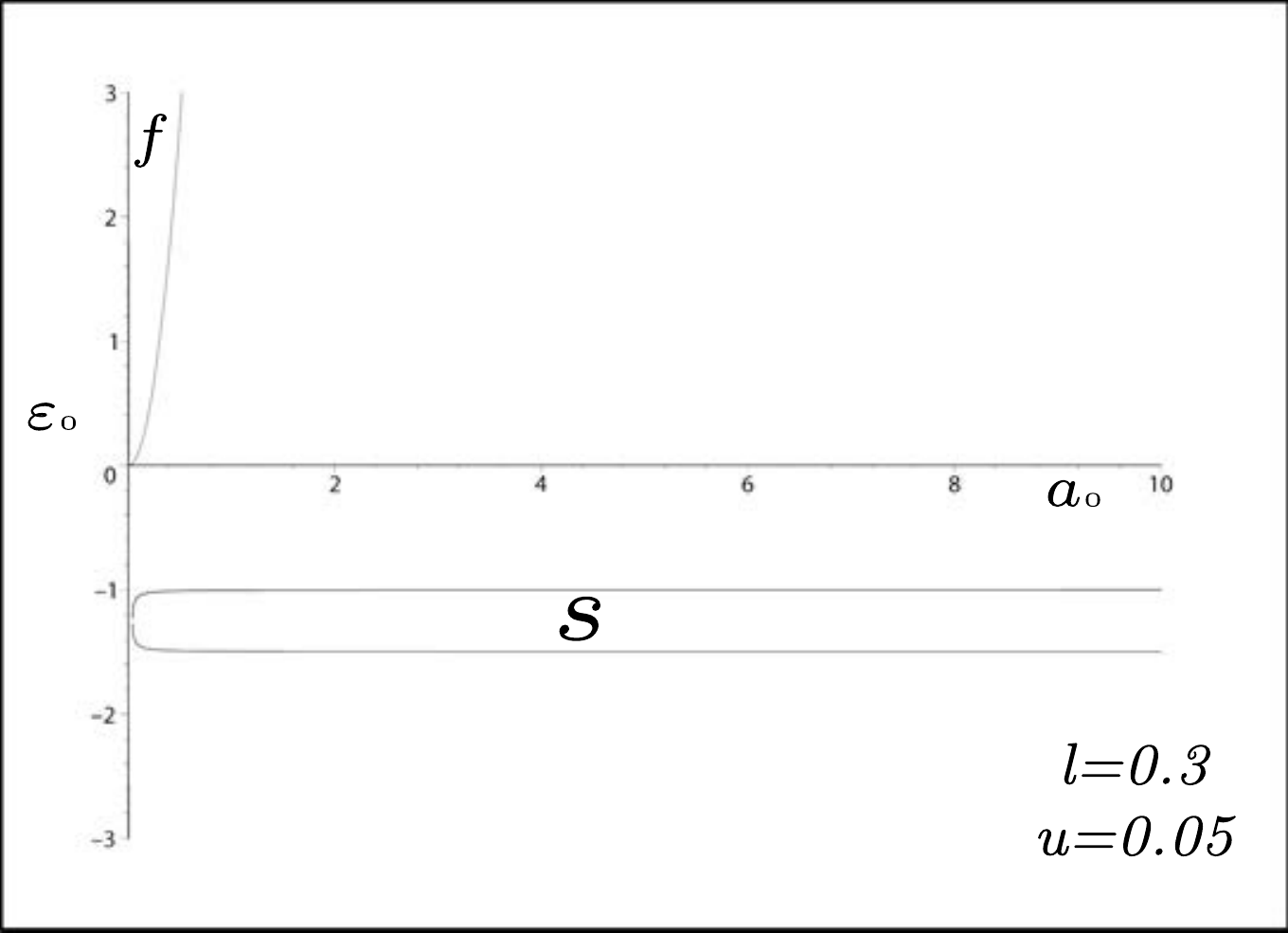} %
\includegraphics[width=0.40\textwidth]{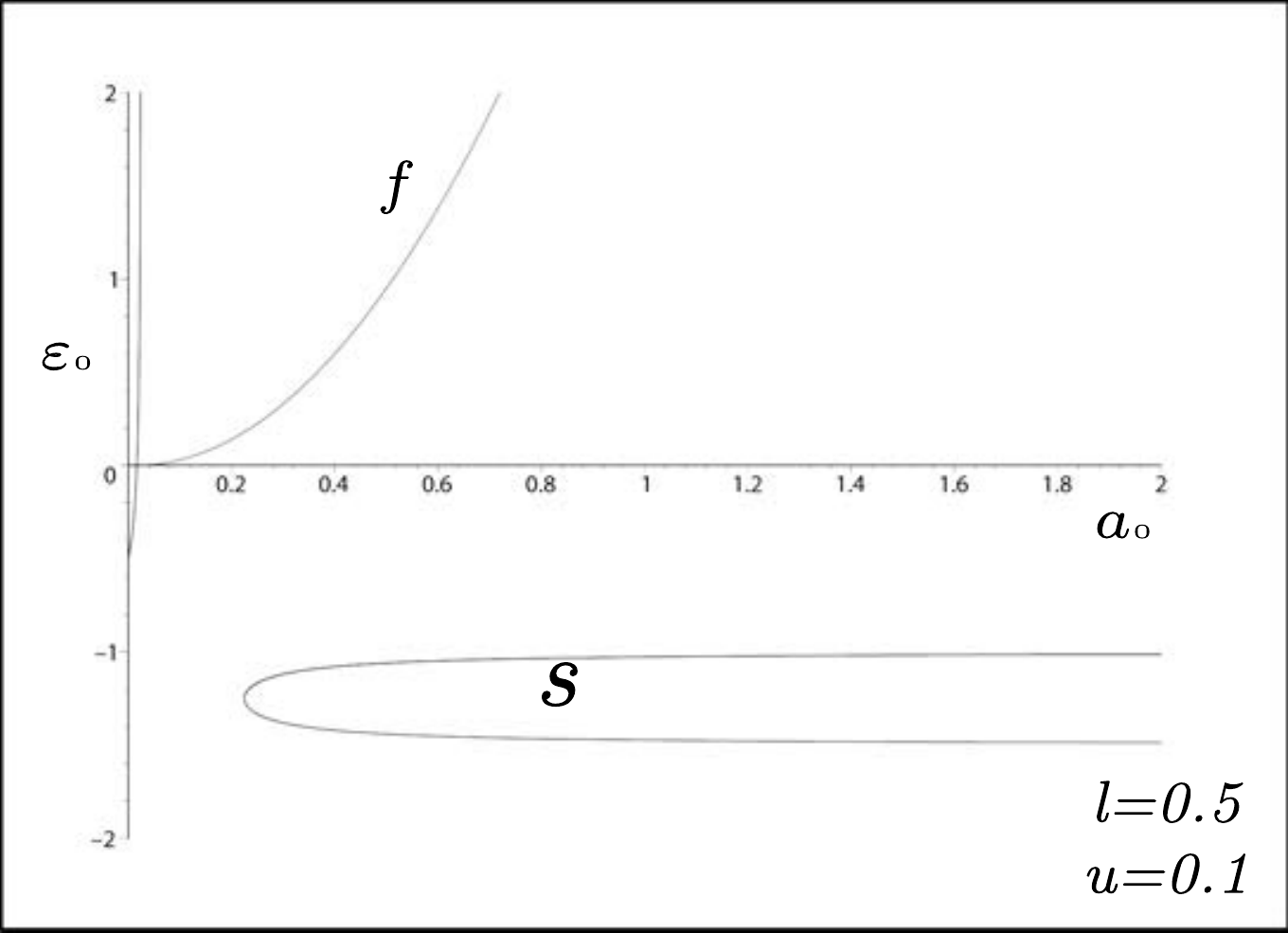} %
\includegraphics[width=0.40\textwidth]{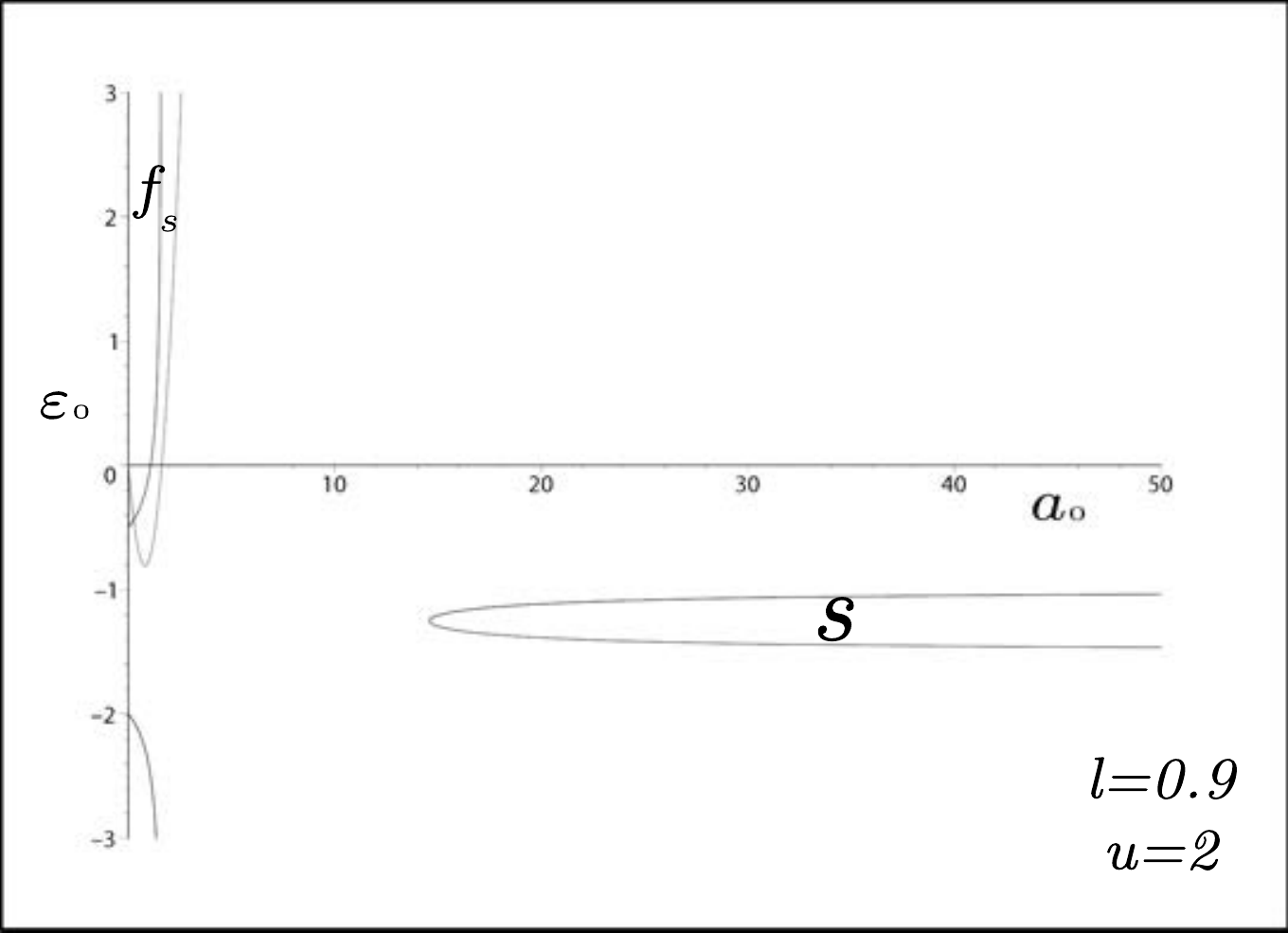}
\end{figure}

\subsection{Stability analysis of TSW via CG}

The equation of state of CG that we considered is given by \cite{cg} 
\begin{equation}
\psi =\varepsilon _{0}(\frac{1}{\sigma }-\frac{1}{\sigma _{0}})+p_{0},
\end{equation}

and one naturally finds 
\begin{equation}
\psi ^{\prime }(\sigma _{0})=\frac{-\varepsilon _{0}}{\sigma _{0}^{2}}.
\end{equation}

After inserting Eq. (39) into Eq. (35), we plot the stability regions for
TSW supported by CG in Fig. (2).

\begin{figure}[h!]
\caption{Stability Regions via the CG}\centering
\includegraphics[width=0.40\textwidth]{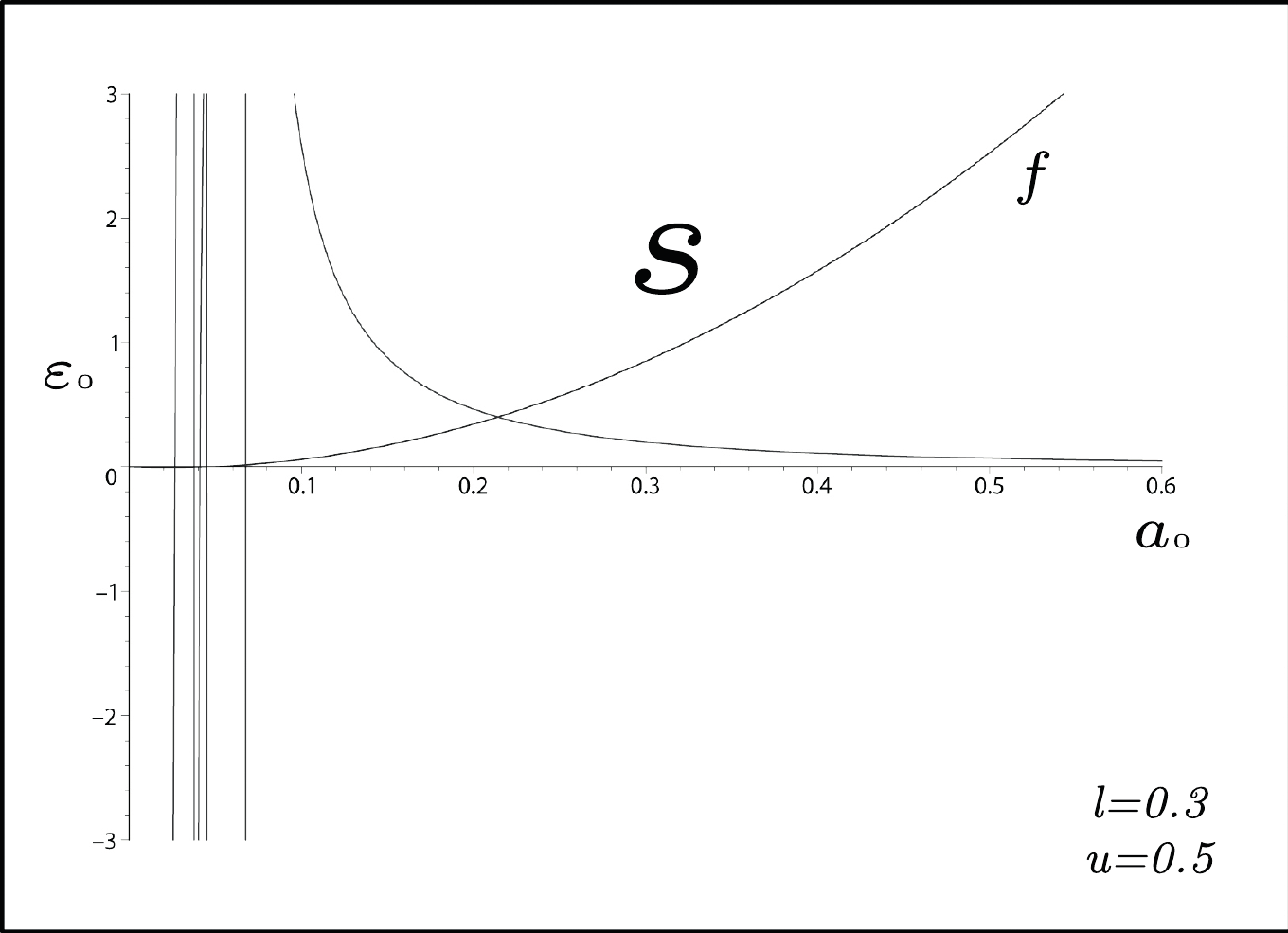} %
\includegraphics[width=0.40\textwidth]{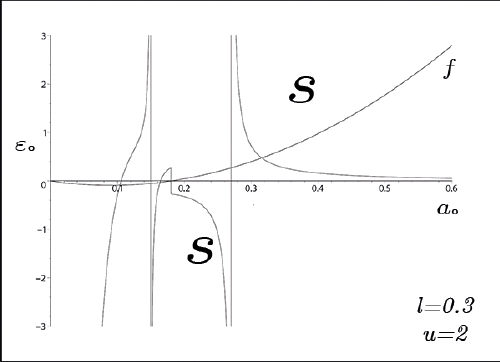} %
\includegraphics[width=0.40\textwidth]{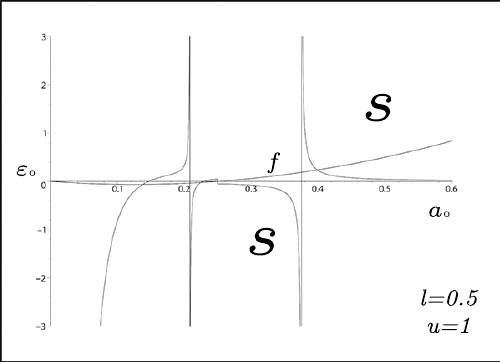} %
\includegraphics[width=0.40\textwidth]{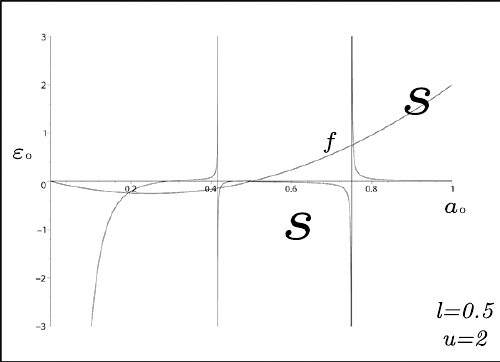}
\end{figure}

\subsection{Stability analysis of TSW via GCG}

By using the equation of state of GCG \cite{gcg}

\begin{equation}
\psi =p_{0}\left( \frac{\sigma _{0}}{\sigma }\right) ^{\varepsilon _{0}},
\end{equation}

and whence 
\begin{equation}
\psi ^{\prime }(\sigma _{0})=-\varepsilon _{0}\frac{p_{0}}{\sigma _{0}},
\end{equation}

Substituting Eq. (41) in Eq. (35), one can illustrate the stability regions
of TSW supported by GCG as seen in Fig. (3). 
\begin{figure}[h]
\caption{Stability Regions via the GCG}\centering
\includegraphics[width=0.40\textwidth]{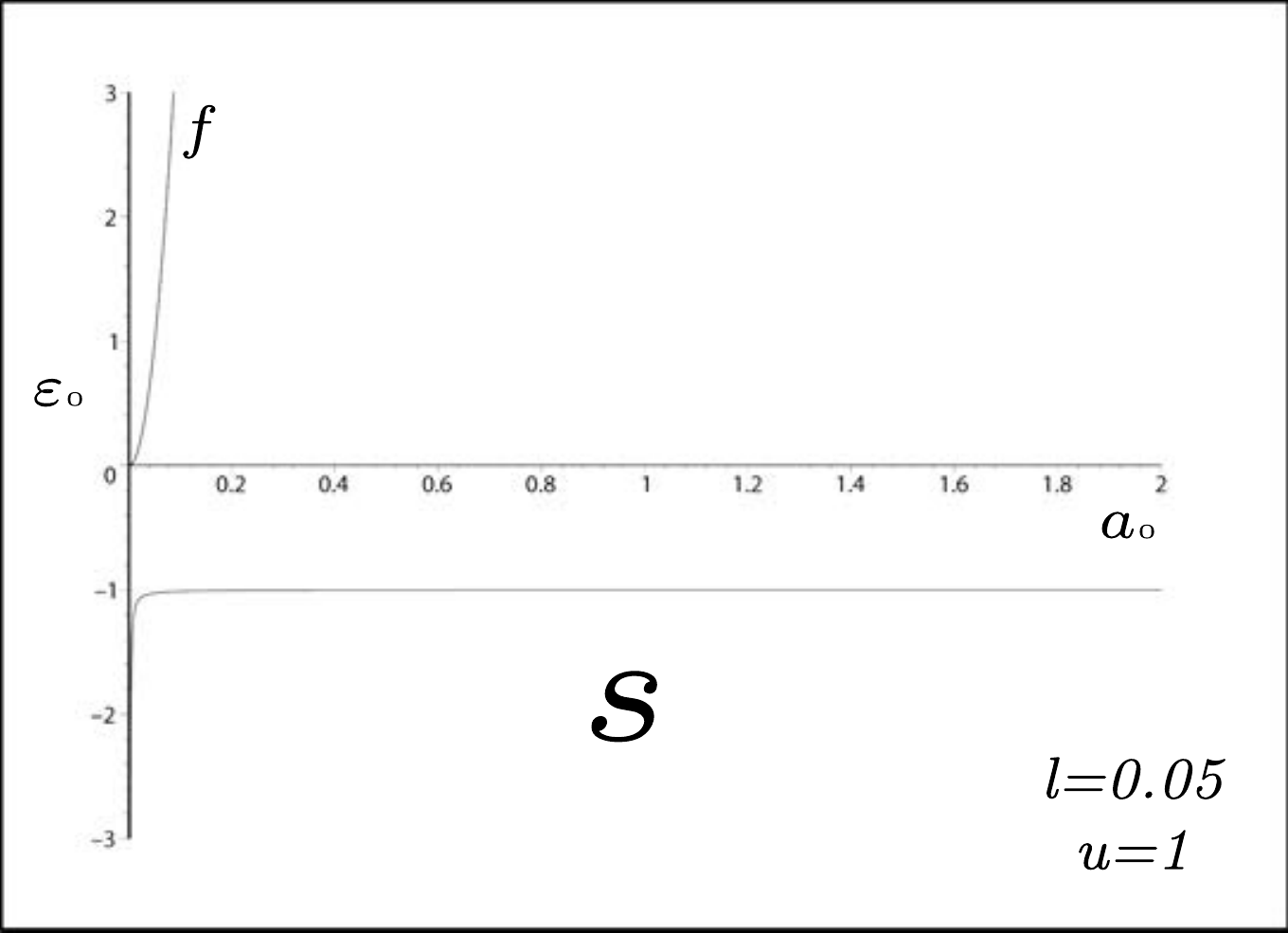} %
\includegraphics[width=0.40\textwidth]{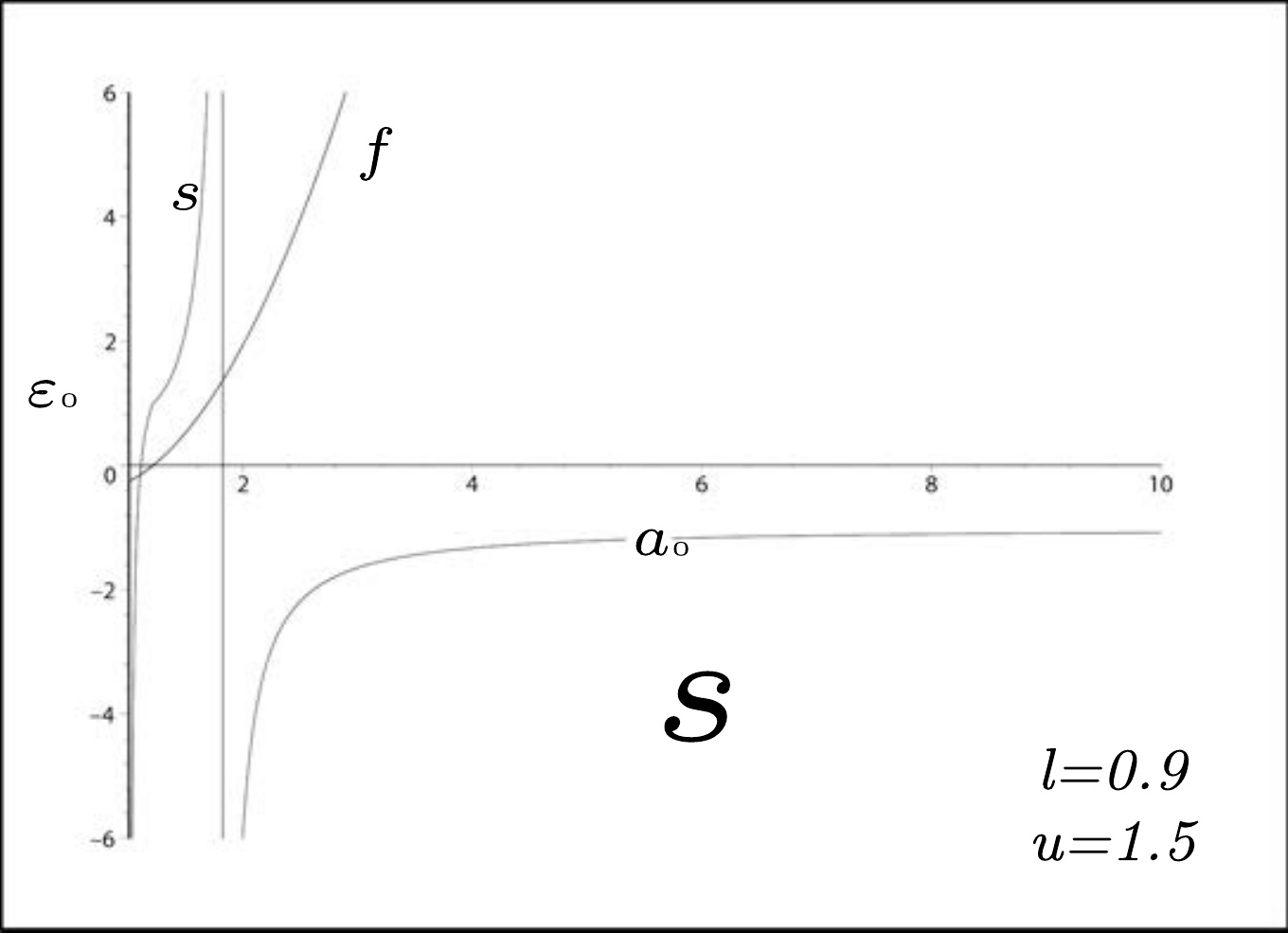} %
\includegraphics[width=0.40\textwidth]{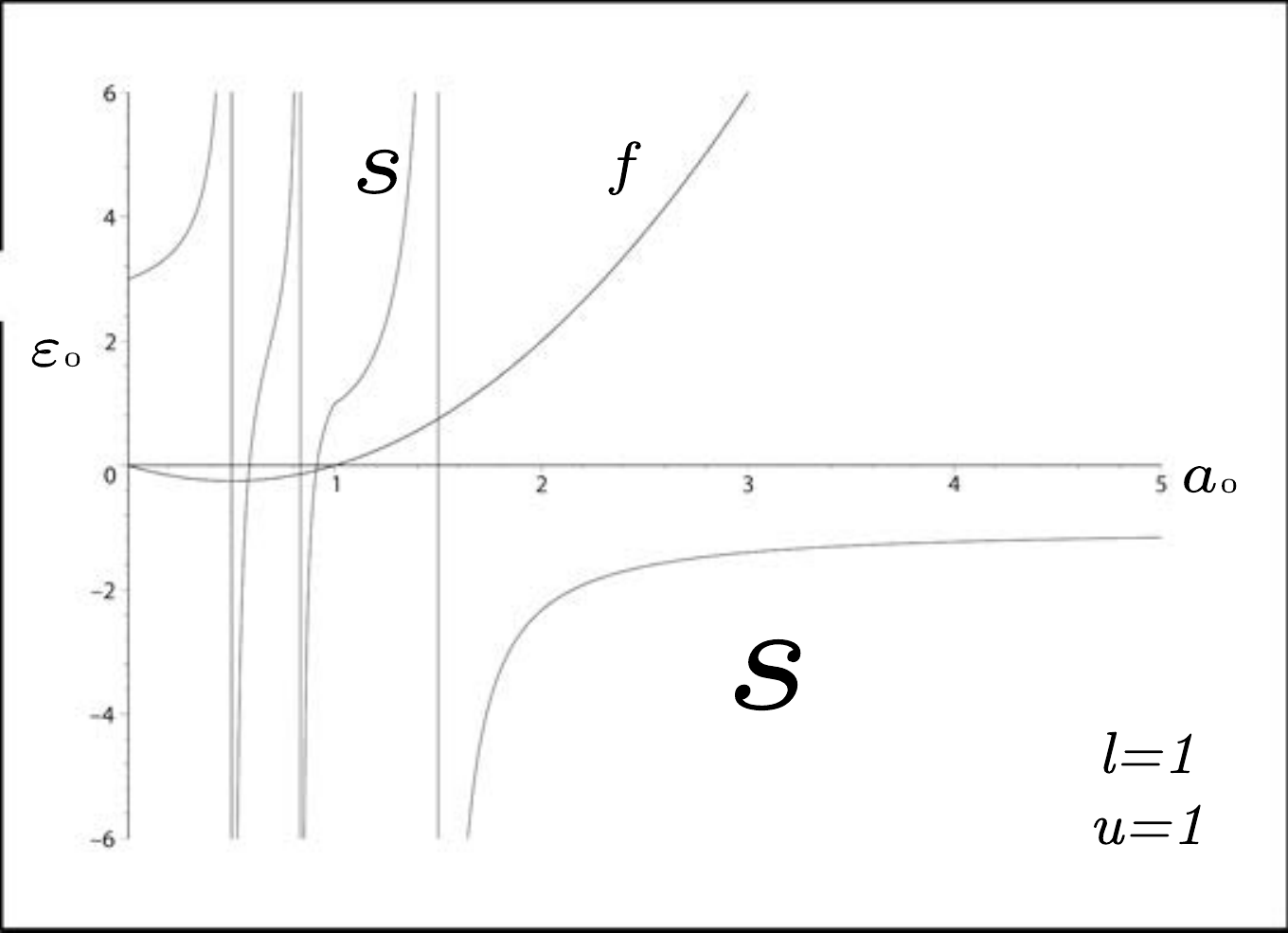} %
\includegraphics[width=0.40\textwidth]{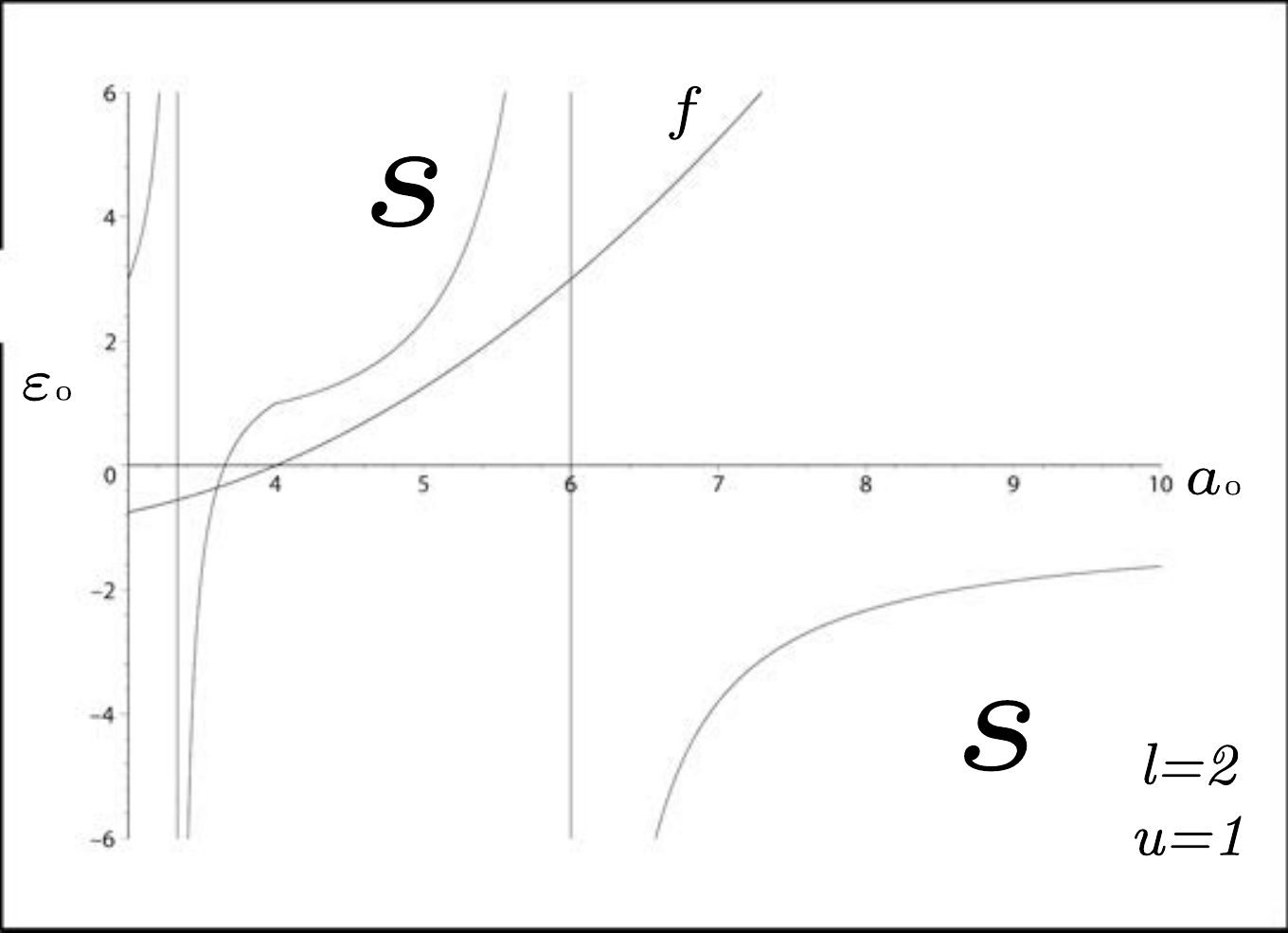}
\end{figure}

\subsection{Stability analysis of TSW via LogG}

In our final example, the equation of state for LogG is selected as follows
(see \cite{ali1}) 
\begin{equation}
\psi =\varepsilon _{0}\ln (\frac{\sigma }{\sigma _{0}})+p_{0},
\end{equation}%
which leads to 
\begin{equation}
\psi ^{\prime }(\sigma _{0})=\frac{\varepsilon _{0}}{\sigma _{0}}.
\end{equation}%
After inserting the above expression into Eq. (35), we show the stability
regions of TSW supported by LogG in Fig. 4.

\begin{figure}[h!]
\caption{Stability Regions via the LogG}\centering
\includegraphics[width=0.40\textwidth]{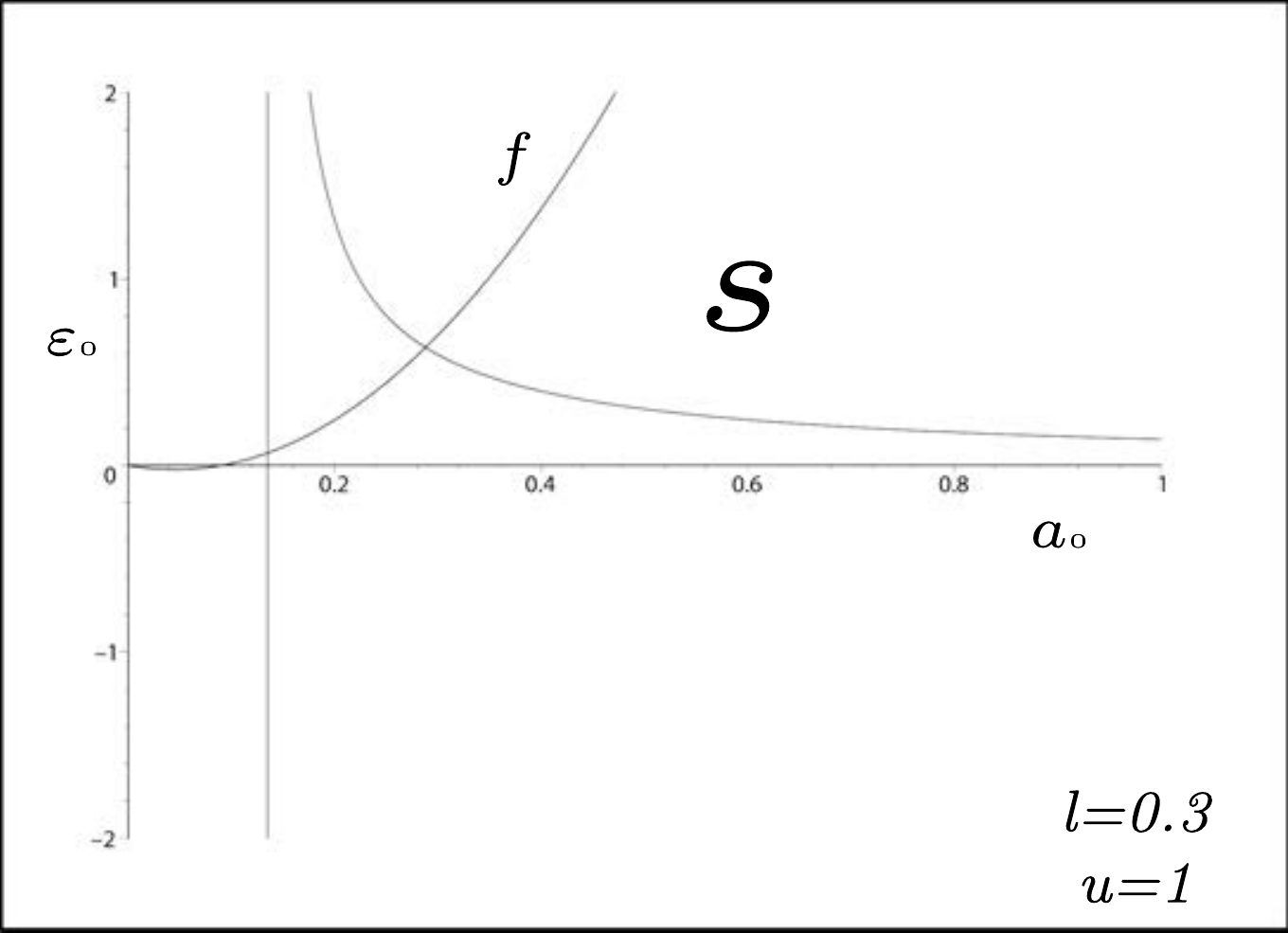} %
\includegraphics[width=0.40\textwidth]{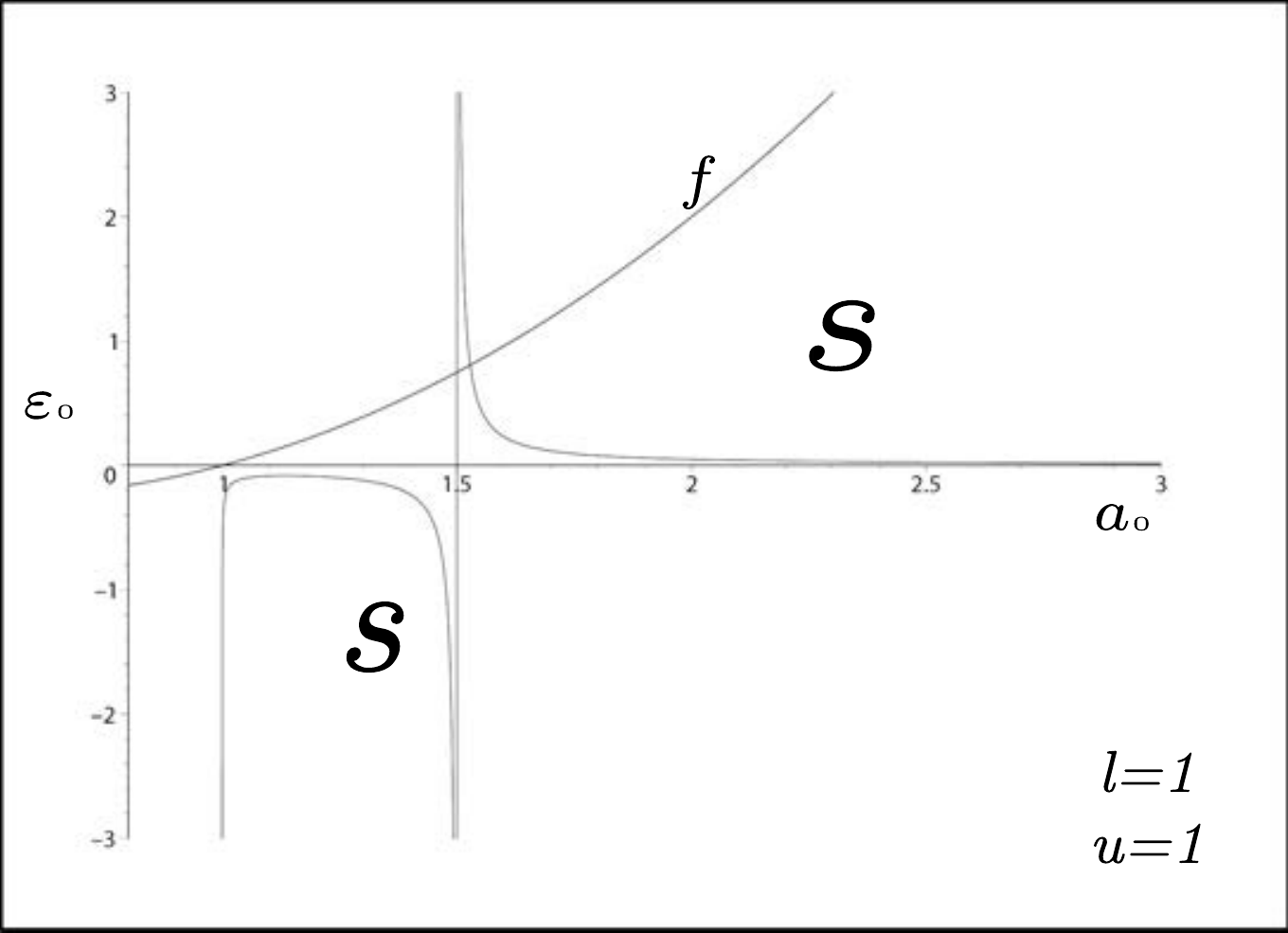} %
\includegraphics[width=0.40\textwidth]{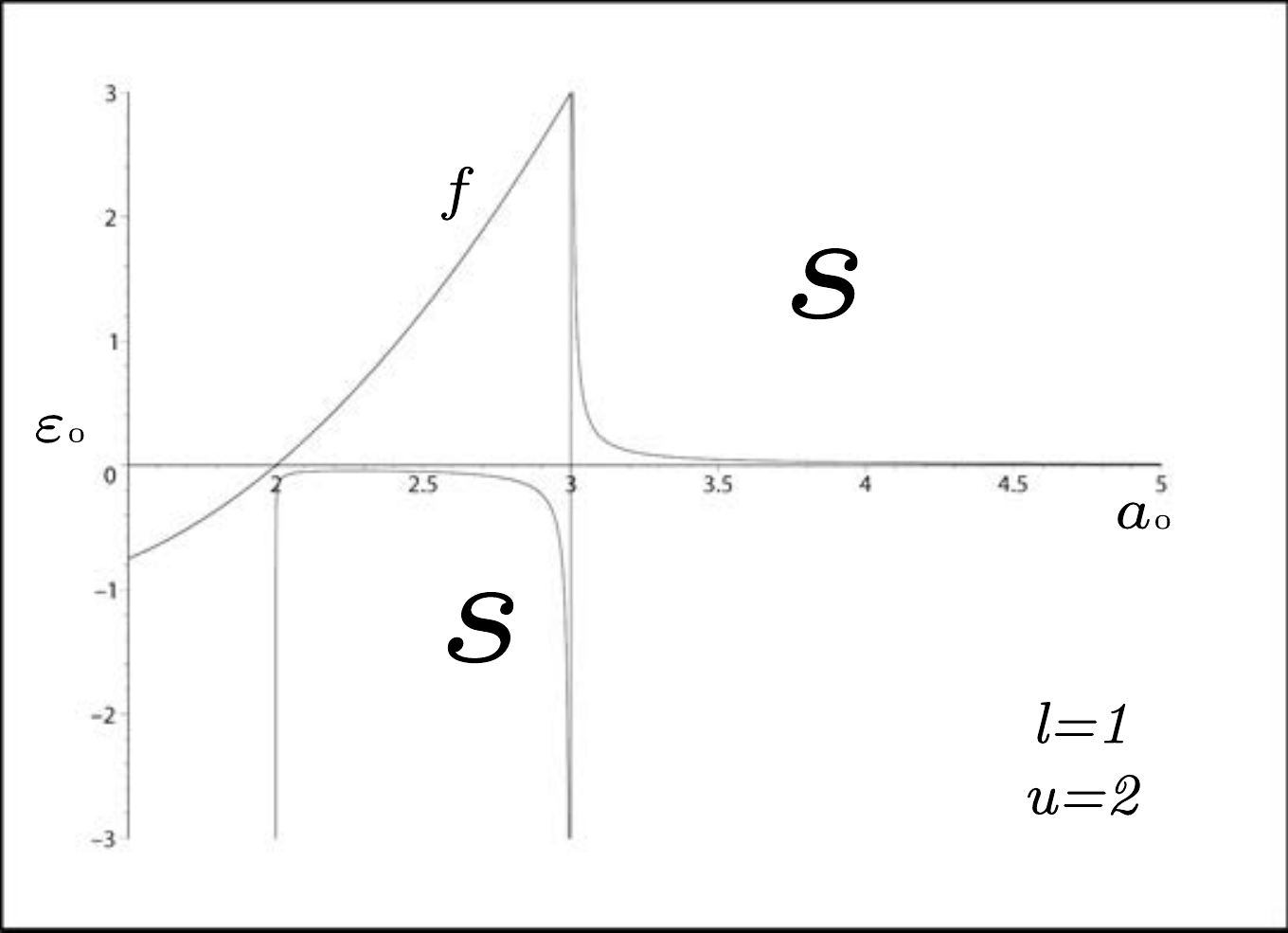} %
\includegraphics[width=0.40\textwidth]{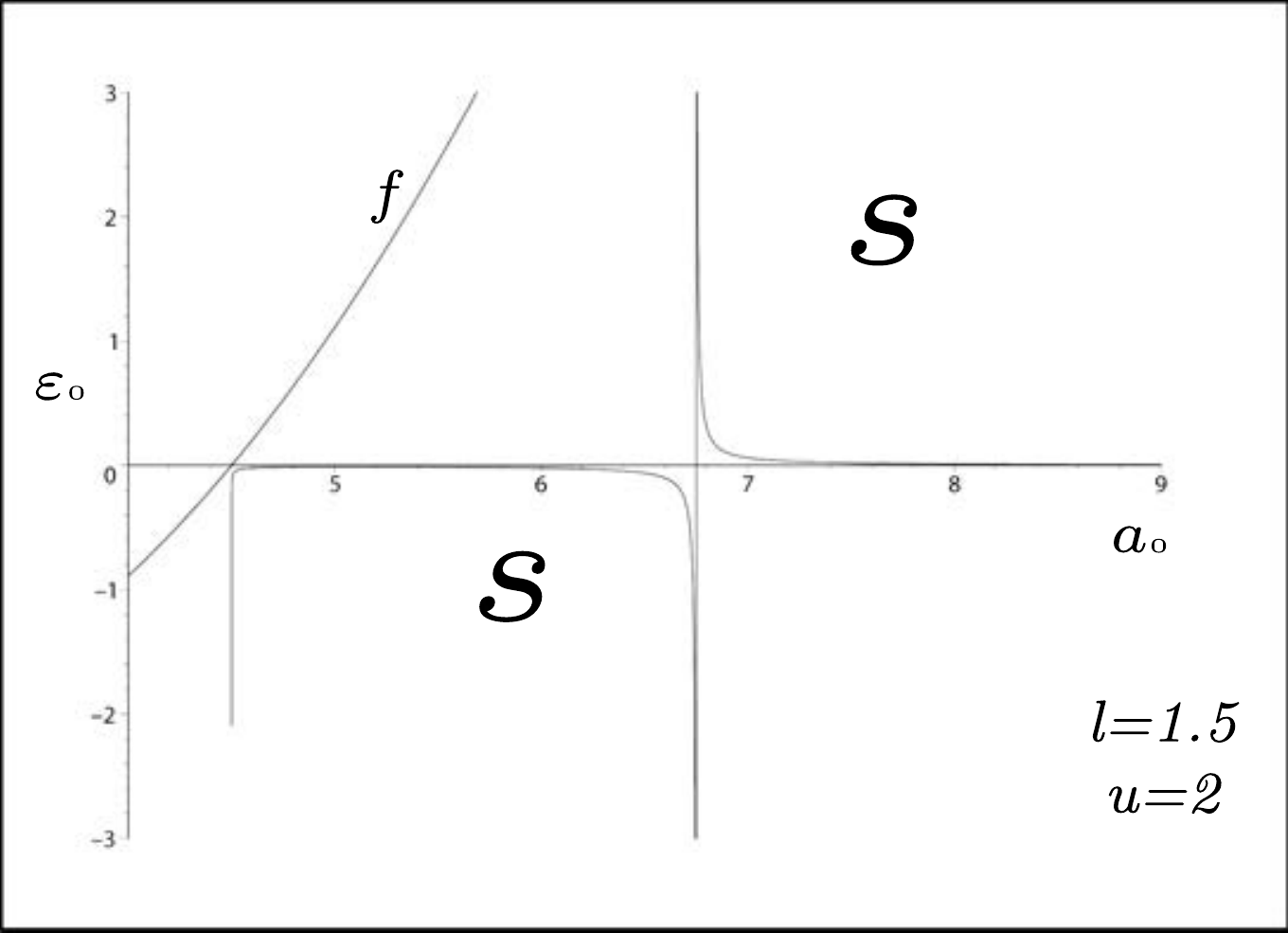}
\end{figure}

\section{Conclusion}

In this paper, we have constructed TSW by gluing two copies of SHBH via the
cut and paste procedure. To this end, we have used the fact that the radius
of throat must be greater than the event horizon of the metric given: ($%
a_{0}>r_{h}$). We have adopted LBG, CG, GCG, and LogG gas equation of states
to the exotic matter locating at the throat. Then, the stability analysis
has become the study of checking positivity of the second derivative of an
effective potential at the throat radius $a_{0}$: $V^{\prime \prime
}(a_{0})\geq 0$. In all cases, we have managed to find the stability regions
in terms of the throat radius $a_{0}$ and constant parameter $\varepsilon
_{0}$, which are associated with the EoS employed. The problem of the
angular perturbation is out of scope for the present paper. That's why we
have only worked on the linear perturbation. However,  angular perturbation
is in our agenda for the extension of this study. This is going to be
studied in the near future.

One of the most trend topics in the theoretical physics is the relationship
between ER and EPR, where ER refers to an Einstein-Rosen bridge (or
wormhole) \cite{er} and EPR, short for Einstein-Podolsky-Rosen \cite{epr},
is another term for the entanglement \cite{suskind,olmo}. In our point of
view, the yet another open problem here is that what is the link between TSW
and EPR? Is it possible to solve exotic matter problem of the TSW by using
the EPR? or vice versa. Also, are there any exotic forces between the EPR
pairs? All of them are still open problems which are awaiting solutions, and
they should be adequately explained. Our next project is to add a small
piece of contribution to this big puzzle..

\section{Acknowledgments}

We are thankful to the editor and anonymous referee for their valuable
comments and suggestions to improve the paper.


\begin{thebibliography}{Eid (2015)}
\bibitem{morris} Morris M.S., Thorne K.S., Am. J. Phys. \textbf{56},395
(1988).

\bibitem{visser1} Hochberg D., Visser M., Phys. Rev. Lett. \textbf{81}, 746
(1998).

\bibitem{visser2} Hochberg D., Molina-Paris C., Visser M., \prd \textbf{59},
044011 (1999).

\bibitem{yurtsever} Morris M.S., Thorne K.S., Yurtsever U., \prl \textbf{61}%
, 1446 (1988).

\bibitem{witt} Friedman J.L., Schleich K., Witt D.M., Phys. Rev. Lett. 
\textbf{71}, 1486 (1995).

\bibitem{nor1} Harko T., Lobo F.S.N., Mak M.K., Sushkov S.V., Phys. Rev. D 
\textbf{87}, 067504 (2013).

\bibitem{nor3} Mazharimousavi S.H., Halilsoy M., Eur.Phys.J. C \textbf{75},
6, 271 (2015).

\bibitem{visser3} Visser M., Nucl. Phys. B \textbf{328 }, 203 (1989).

\bibitem{israel} Israel W., II Nuovo Cimento B \textbf{10},44 (1966).

\bibitem{visser4} Poisson E., Visser M., Phys. Rev. D \textbf{52}, 7318
(1995).

\bibitem{banerjee2} Banerjee A., Int. J. Theor. Phys. \textbf{52}, 2943
(2013).

\bibitem{banerjee7} Banerjee A, Rahaman F., Chattopadhyay S., Banerjee S.,
Int. J. Theor. Phys. \textbf{52}, 3188 (2013).

\bibitem{banerjee3} Banerjee A, Rahaman F., Jotania K., Sharma R., Rahaman
M., Astrophys. Space Sci. \textbf{355}, 353 (2015).

\bibitem{banerjee6} Bhar P., Banerjee A, I Int.J.Mod.Phys. D. \textbf{24},
05, 1550034 (2015).

\bibitem{sime1} Bejarano C., Eiroa E.F., Simeone C., Eur. Phys. J. C \textbf{%
74}, 3015 (2014).

\bibitem{darabi} Darabi F.,Theor.Math.Phys. \textbf{173}, 1734-1742 (2012).

\bibitem{nor2} Mazharimousavi S.H., Halilsoy M., Eur. Phys. J. C \textbf{74}%
, 9 , 3073 (2014).

\bibitem{lemos1} Dias G.A.S., Lemos J.P.S., Phys. Rev. D \textbf{82}, 084023
(2010).

\bibitem[Eid (2015)]{eid} Eid A., New Astronomy \textbf{39}, 72 (2015).

\bibitem{lobo} Garcia N.M., Lobo F.S.N., Visser M., Phys. Rev. D \textbf{86}%
, 044026 (2012).

\bibitem{ali1} Halilsoy M., Ovgun A., Mazharimousavi S. H., Eur. Phys. J. C 
\textbf{74}, 2796 (2014).

\bibitem{Kas} Kashargin P.E., Sushkov S.V., Grav.Cosmol. \textbf{17},
119-125 (2011).

\bibitem{kuf} Kuhfittig P.K.F., Adv. High Energy Phys. \textbf{2012}, 462493
(2012).

\bibitem{camera} La Camera M., Mod. Phys. Lett. A \textbf{26}, 857 (2011).

\bibitem{banerjee8} Rahaman F., Banerjee A., Int. J. Theor. Phys. \textbf{51}%
, 901 (2012).

\bibitem{banerjee} Rahaman F., Banerjee A., Radinschi I., Int. J. Theor.
Phys. \textbf{51}, 1680 (2012).

\bibitem{banerjee9} Rahaman F., Kuhfittig P.K.F., Kalam M., Usmani A.A., Ray
S., Class.Quant.Grav. \textbf{28}, 155021 (2011).

\bibitem{banerjee10} Rahaman F., Rahman K.A., Rakib Sk.A., Kuhfittig P.K.F.,
Int.J.Theor.Phys. \textbf{49}, 2364 (2010).

\bibitem{sharif1} Sharif M., Azam M., Eur. Phys. J. C \textbf{73}, 2407
(2013).

\bibitem{sharif2} Sharif M., Azam M., JCAP \textbf{1304} , 023 (2013).

\bibitem{sharif3} Sharif M., Azam M., JCAP \textbf{1305}, 025 (2013).

\bibitem{sharif4} Sharif M., Mumtaz S., Astrophys. Space Sci. \textbf{352},
729 (2014).

\bibitem{sharif5} Sharif M., Azam M., Phys. Lett. A \textbf{378 }, 2737
(2014).

\bibitem{MH} Mazharimousavi S.H., Halilsoy M., Eur. Phys. J. Plus \textbf{130%
}, 158 (2015).

\bibitem{sakalli0} Mazharimousavi S.H., Halilsoy M., Sakalli I., Gurtug O.,
Class. Quantum Gravity \textbf{27}, 105005 (2010).

\bibitem{sakalli1} Pasaoglu H., Sakalli I., Int. J. Theor. Phys. \textbf{48}%
, 3517 (2009).

\bibitem{Clem1} Clement G., Fabris J.C., Marques G.T., Phys.Lett.B \textbf{%
651}, 54 (2007).

\bibitem{sakalli2} Sakalli I., Halilsoy M., Pasaoglu H., Astrophys. Space
Sci. \textbf{340}, 155 (2012).

\bibitem{ali2} Sakalli I., Ovgun A., Europhys. Lett. \textbf{110}, 10008
(2015).

\bibitem{ali22} Sakalli I., Ovgun A., Astrophys. Space Sci. \textbf{359}, 32
(2015).

\bibitem{sakalli11} Sakalli I., Halilsoy M., Pasaoglu H., Int. J. Theor.
Phys. \textbf{50}, 3212 (2011).

\bibitem{sakalli12} Sakalli I., Int. J. Theor. Phys. \textbf{50}, 2426
(2011).

\bibitem{darmois} Darmois G., M\'{e}morial des Sciences Math\'{e}matiques,
Fascicule XXV (Gauthier-Villars, Paris, 1927).

\bibitem{lan1} Lanczos K., Ann. Phys. (Leipzig) \textbf{379}, 518 (1924).

\bibitem{lake} Musgrave P., Lake K., Class. Quantum Gravity \textbf{13},
1885 (1996).

\bibitem{nandi} Nandi K.K., Zhang Y.Z., Kumar K.B.V., Phys. Rev. D \textbf{70%
}, 127503 (2004).

\bibitem{bro1} Bronnikov K. A., Lipatova L. N., Novikov I. D., Shatskiy A.
A., Grav. Cosmol. \textbf{19}, 269 (2013).

\bibitem{igor} Novikov I., Shatskiy A., J.Exp.Theor.Phys, \textbf{141}, 5
(2012).

\bibitem{kuf2} Kuhfittig P.K.F., Ann. Phys. \textbf{355}, 115 (2015).

\bibitem{varela} Varela V., Phys. Rev. D. \textbf{92}, 044002 (2015).

\bibitem{cg} Eiroa E.F., Simeone C., Phys. Rev. D \textbf{76}, 024021 (2007).

\bibitem{cg1} Lobo F.S.N., Phys. Rev. D \textbf{73}, 064028 (2006).

\bibitem{gcg} Gorini V., Moschella U., Kamenshchik A. Y.,bPasquier V., and
A. A. Starobinsky,Phys. Rev. D \textbf{78}, 064064 (2008).

\bibitem{er} Einstein A., Rosen N., Phys. Rev.\textbf{\ 48}, 73 (1935).

\bibitem{epr} Einstein A., Podolsky B., Rosen N., Phys. Rev.\textbf{47},777
(1935).

\bibitem{olmo} Lobo F.S.N., Olmo G.J., Rubiera-Garcia D., Eur. Phys. J. C 
\textbf{74}, 2924 (2014).

\bibitem{suskind} Maldacena J., Susskind L., Fortschr. Phys. \textbf{61},
781 (2013).
\end{thebibliography}
\end{document}